\documentclass[onecolumn,showpacs,prl]{revtex4}

\usepackage{graphicx}
\usepackage{dcolumn}
\usepackage{bm}
\usepackage{amsfonts}
\usepackage{amssymb}

\usepackage{epsfig}
\usepackage{amsmath}

\begin{document}

\title{Frequency dependent polarizability of small metallic grains}
\author{A.~A.~Zharov and I.~S.~Beloborodov}

\address{Department of Physics and Astronomy, California State University Northridge, Northridge, CA 91330, USA}

\begin{abstract}
We study the dynamic electronic polarizability of a single nano-scale spherical metallic grain using quantum mechanical approach. We introduce the model for interacting electrons bound in the grain allowing us numerically to calculate the frequency dependence of the polarizability of grains of different sizes. We show that within this model the main resonance peak corresponding to the surface plasmon mode is blue-shifted and some minor secondary resonances above and below the main peak exist. We study the behavior of blue shift as a function of grain size and compare our findings with the classical polarizability and with other results in the literature.
\end{abstract}


\maketitle

\section{Introduction}
Great efforts in contemporary materials science research focus on properties of granular materials~\cite{Beloborodov07}. The interest is motivated by the fact that granular arrays can be treated as artificial solids with programmable electronic properties. The ease of adjusting electronic properties of granular materials is one of their most attractive assets for fundamental studies of disordered solids and for targeted applications in nanotechnology.  The parameters of granular materials in many ways are determined by the properties of their building blocks - grains.

In this paper we study the dynamic electronic polarizability of a single nano-scale spherical metallic grain. This research is important because on one hand confinement effects can drastically modify the dynamic polarizability of metallic grains and on the other hand the fundamental question related to the description of crossover between microscopic and macroscopic behavior of the grains exists.

The electronic polarizability was first quantum mechanically studied in Ref.~\cite{Gorkov}. However, the influence of screening effects on
static polarizability was first investigated in Ref.~\cite{Rice}. Later the dynamic polarizability of very small metallic particles was considered within the jellium-background model~\cite{Ekardt,SpherJ,Beck,Koch,Weick}.
This model was also used to study the electronic structure and polarizability of metallic nanoshells~\cite{Prodan}.
The electronic polarizability was also studied in applications for optical properties
of small metallic grains~\cite{Ruppin,Halperin,Wood,Genzel2}. More recently the nonperturbative supersymmetry technique was used to consider the  effects of disorder on properties of electronic polarizability~\cite{Efetov}.

Here we study the dynamic electronic polarizability of a single metallic grain using quantum mechanical approach. We introduce the model for interacting electrons bound in the grain allowing us numerically to calculate the frequency dependence of the polarizability of grains of different sizes.

\section{Main Results} 
\begin{figure}
\includegraphics[width=4in]{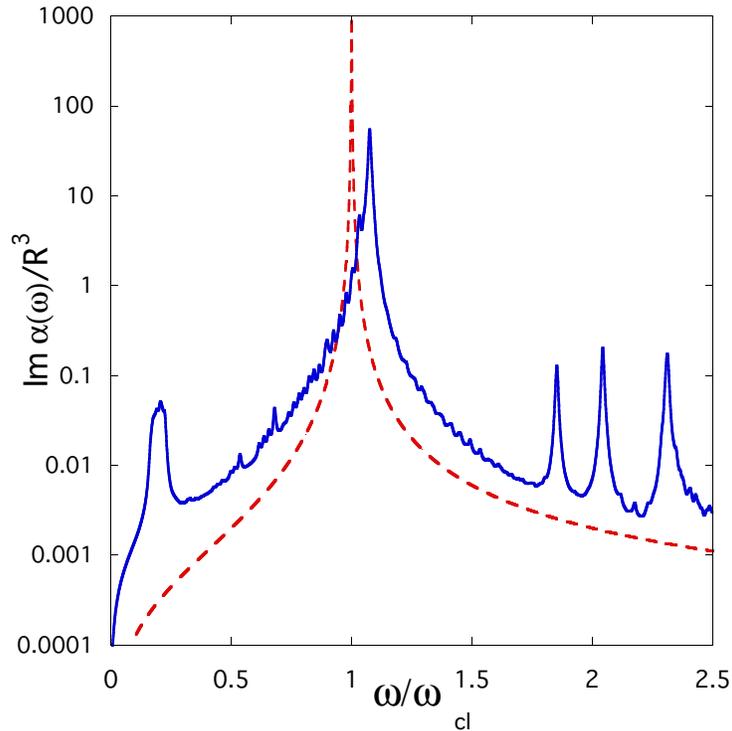}
\caption{(Color online) Imaginary part of dynamic electronic polarizability $\alpha(\omega)$ of the metallic grain (in units of $R^3$, $R$ being the radius of a grain) vs dimensionless frequency $\omega/\omega_{cl}$ with  $\omega_{cl}=\omega_p/\sqrt{3}$ being the classical resonance frequency, where $\omega_p$ is the plasma resonance frequency. The blue solid line represents the numerical solution for polarizability $\alpha(\omega)$ corresponding to the number of electrons $N = 16304$, $R= 5.28$ nm($\alpha_F=50$, see Eq.~(\ref{destribution})). The red dashed line represents the classical Drude polarizability $\alpha_{cl}(\omega)$, Eq.~(\ref{imcl}), for comparison.}
\label{fig50}
\end{figure}

\begin{figure}
\includegraphics[width=4in]{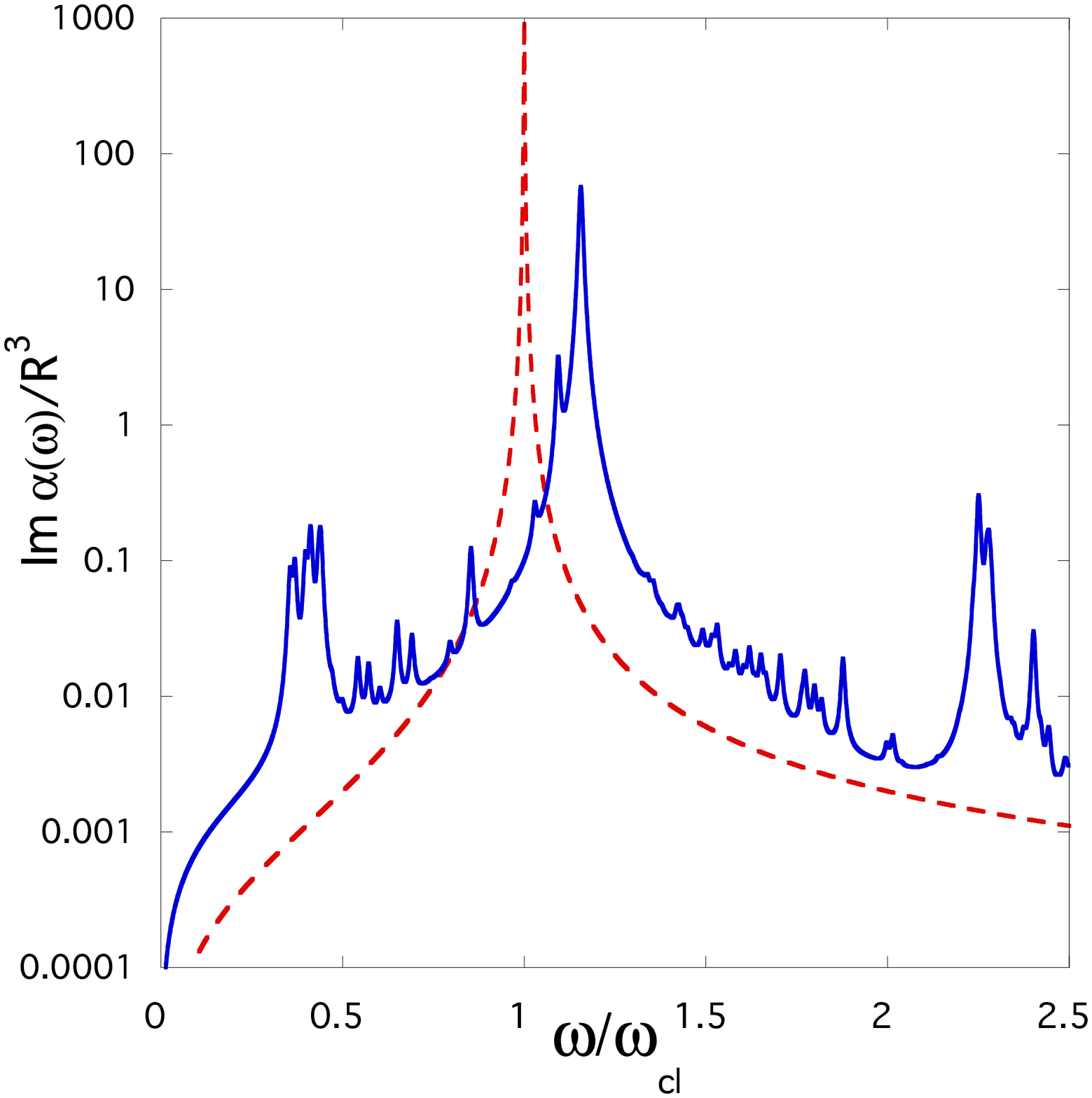}
\caption{(Color online) Imaginary part of dynamic electronic polarizability $\alpha(\omega)$ of the metallic grain (in the units of $R^3$, $R$ being the radius of a grain) vs dimensionless frequency $\omega/\omega_{cl}$ with  $\omega_{cl}=\omega_p/\sqrt{3}$ being the classical resonance frequency.  The blue solid line represents the numerically solution for polarizability $\alpha(\omega)$ corresponding to the number of electrons $N = 1956$, $R= 2.6$ nm ($\alpha_F=25$, see Eq.~(\ref{destribution})). The red dashed line represents the classical Drude polarizability $\alpha_{cl}(\omega)$, Eq.~(\ref{imcl}), for comparison.}
\label{fig25}
\end{figure}

We numerically calculate the dynamic polarizability $\alpha(\omega)$ of grains of
different sizes. The experimentally measurable quantity is the photoabsorption cross section $\sigma(\omega)$. It is related
to the imaginary part of the polarizability $\alpha(\omega)$ as follows
\begin{equation}
\label{sigma}
\sigma(\omega)=\frac{4\pi\omega}{c}\textrm{Im}\, {\alpha(\omega)},
\end{equation}
where $c$ is the speed of light.
The photoabsorption cross section $\sigma(\omega)$ is the main result of this work. It is plotted in Figs.~\ref{fig50} and \ref{fig25} for two grains of different sizes (the number of electrons $N=16304$ and $N=1956$, corresponding to the grain sizes of 5.28 nm and 2.6 nm, respectively).

Our calculations were done at zero temperature. Therefore, due to Fermi-Dirac statistics all electron states are occupied below the Fermi level and empty above it. As a result, the total number of electrons $N$, that determines the size of the grain, is related to the Fermi energy level. This number can be found by calculating the total number of  electron states below the Fermi level. The Fermi energy can be parameterized by dimensionless quantity $\alpha_F$ as $E_F=(\hbar^2/2mR^2)\alpha_F^2$, where $R$ is a grain radius. In order to estimate the grain size we need to know the total number of electrons $N$ and the electron density $r_s$. In terms of these parameters the grain size is given by $R/a_B=N^{1/3}r_s$, where $a_B$ is Bohr radius. All grain size estimates were done for sodium ($r_s=3.93$).


The behavior of dynamic polarizability $\alpha(\omega)$ presented in Figs.~\ref{fig50} and \ref{fig25} revels several features: i) the main resonance peak is slightly shifted (with respect to the classical result) to higher frequency; ii) some smaller secondary resonances above and below the main resonance frequency exist. Both of these features have a tendency to decrease as the grain size increases which is consistent with the fact that macroscopic grain has a fully classical behavior.

The main resonance peak in Figs.~\ref{fig50} and \ref{fig25} corresponds to the surface plasmon excitation.
The blue-shift of the main resonance peak in comparison with the classical Mie resonance (the red dash line in Figs.~\ref{fig50} and \ref{fig25}) can be understood as follows: In a quantum mechanical problem due to Fermi-Dirac statistics electron transitions may occur only from an occupied state to an empty state. For higher external frequency the total number of unrestricted electron transitions becomes higher. Therefore in a quantum mechanical problem if the external frequency equals to the classical plasmon frequency, $\omega_{cl}=\omega_p/\sqrt{3}$ in Figs.~\ref{fig50} and \ref{fig25}, the resonance amplitude is suppressed because of a lack of possible electron transitions. Hence the main resonance peak has a tendency to move to a higher frequency.

The surface plasmon excitations were also studied quantum mechanically in Ref.~\cite{Ovchinnikov}, and using extended classical Mie theory in Ref.~\cite{Ruppin}, where it was shown that the frequency of a surface plasmon for very small grains is blue-shifted. This is in agreement with our result. Several experimental works have also observed the blue-shifted surface plasmon resonance peak~\cite{Duthler,Genzel3}.

A different result was obtained in Refs.~\cite{Ekardt,Weick}, where the main resonance peak was red-shifted. This discrepancy may be explained by the difference between the models used. In particular, in Ref.~\cite{Ekardt} the electronic polarizability was studied within the spherical jellium-background model. The red-shift in this model is explained by the electron spill-out effect. Here we use "electron in a box" approximation. In this model the spill-out effect is prevented by the infinitely deep potential well.

The width of the plasmon resonance in Figs.~\ref{fig50} and \ref{fig25} is determined by the artificially introduced imaginary part of the frequency $\omega$ to avoid infinite values of polarizability $\alpha(\omega)$ at the resonances. This width does not correspond to a real width of the photoabsorbtion cross section $\sigma(\omega)$ in Eq.~(\ref{sigma}). We discuss this issue in more details in the "solution" section below.

The secondary resonances shown in Figs.~\ref{fig50} and \ref{fig25} are significantly lower than the main peak. These resonances can be divided into two groups: i) low frequency single particle excitations and ii) high frequency resonances corresponding to the transitions between the states with higher orbital quantum numbers. Both of these types of excitations become less pronounced with increasing grain size. We notice that at low energies the transitions may occur only in the immediate vicinity to the Fermi surface. Therefore the low frequency peaks in Figs.~\ref{fig50} and \ref{fig25} correspond to the resonant transitions between the states which are located just below and just above the Fermi surface.

We now comment on the influence of disorder in the grain on the resonance peaks. First, we notice, that our consideration is based on the fact, that the disorder is weak. Thus, the conducting electrons are delocalized in our model. The characteristic energy scale for disorder is the inverse scattering time $1/\tau$. In the limit of weak disorder the electron propagation within the grain is almost ballistic and therefore $1/\tau$ is of the order of Thouless energy, $1/\tau \sim E_{Th}$. However, both of these energy scales are smaller than the plasma frequency, $\omega_p$, which is the characteristic scale for the resonances. Therefore, we expect that for energies less than $\omega_p$ the disorder is irrelevant and can not modify the picture qualitatively.

However, the presence of disorder may lead to the broadening and suppression of all of the resonance peaks. The broadening effects are beyond our present consideration. The disorder caused suppression might make the secondary resonances hard to detect, as they are small compared to the main resonance peak.

We also mention that our consideration can be generalized for the description of slightly non-spherical grains. In this case the non ideal grain shape can be studied within the perturbation theory. It will result in the changes of the energy levels structure. In particular, the degenerate energy levels will be split. That will slightly affect the positions of the resonance peaks.

To conclude the analysis of the polarizabilities we compare our findings for $\alpha(\omega)$ with the known result for classical polarizability $\alpha_{cl}(\omega)$,~\cite{Mie1908}
\begin{equation}
\label{classical}
\alpha_{cl}(\omega)=R^3\, \frac{\epsilon(\omega)-1}{\epsilon(\omega)+2},
\end{equation}
where $R$ is the radius of a grain and $\epsilon(\omega)$ is the grain dielectric constant; within the Drude model $\epsilon(\omega) = 1-\frac{\omega_p^2}{\omega(\omega+i\delta)}$,
where $\omega_p = ((4/3)\pi\nu)^{1/2} e v_F$ is the plasma resonance frequency and $\delta$ is the damping factor. The imaginary part of the polarizability $\alpha_{cl}(\omega)$ in Eq.~(\ref{classical}) is given by
\begin{equation}
\label{imcl}
\textrm{Im}\, \alpha_{cl}(\omega')=R^3\frac{\omega'\gamma}{(1-\omega')^2+\gamma^2\omega'^2},
\end{equation}
where we introduce the dimensionless parameter $\gamma=\delta/(\omega_p/\sqrt{3})$ and the dimensionless frequency
$\omega' = \omega/(\omega_p/\sqrt{3})$. The right hand side of Eq.~(\ref{imcl})
has a single resonance peak at frequency $\omega'=1$ (or $\omega=\omega_p/\sqrt{3}$) corresponding to the surface plasmon mode excitation.
We plotted $\textrm{Im}\, \alpha_{cl}(\omega)$ in Figs.~\ref{fig50} and \ref{fig25} for comparison with our results.

\begin{figure}
\includegraphics[width=4in]{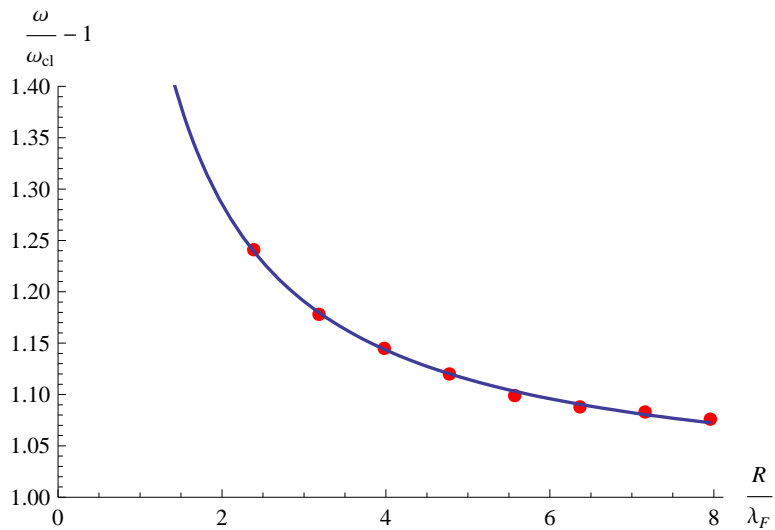}
\caption{(Color online) Blue shift of the surface plasmon frequency (in the units of the classic surface plasmon frequency $\omega_{cl}=\omega_p/\sqrt{3}$, with $\omega_p$ being the plasma frequency) as a function of the dimensionless grain radius, $R/\lambda_F$, where $R$ is the grain radius and $\lambda_F$ is a Fermi wavelength. Red dots represent the numerical calculations. The blue solid line represents the fitting of the numerical calculations with the function $(R/\lambda_F)^{-1}$}
\label{fig_size}
\end{figure}

The last important question to discuss is the dependence of the surface plasmon frequency on the grain size. This dependence is shown in Fig.~\ref{fig_size}. The calculated red dots were then fitted with the function that scales as $(R/\lambda_F)^{-1}$. On can clearly see that the blue shift has the tendency to decrease as the grain size increases.

Now we turn to the description of our model and the derivation of electronic polarizability $\alpha(\omega)$ in Eq.~(\ref{sigma}).

\section{The Model} A single spherical metallic grain is described by the following Hamiltonian
\begin{subequations}
\begin{equation}
\label{hamiltonian}
\hat{H}=\hat{H}_0+\hat{H}_{int}.
\end{equation}
Here
\begin{equation}
\label{h0}
\hat{H}_0 = \int d^3r \psi^{\dag}(\textbf{r}) \left( - \frac{\nabla^2}{2m} + U(\textbf{r})\right) \psi(\textbf{r}),
\end{equation}
is the Hamiltonian of the non-interacting electrons trapped in the infinitely deep spherical potential well $U(\textbf{r})$:
\begin{equation}\label{potential}
U(\textbf{r})=\left\{
\begin{array}{lr}
0, \hspace{0.3cm} |\textbf{r}|<R \\
\infty, \hspace{0.3cm} |\textbf{r}|>R
\end{array}
\right.
\end{equation}
This form of the potential allows us simple analytical expansion in eigenfunctions of the Hamiltonian $\hat{H}_0$, which in its turn allows us to use the many-body approach.
The boundary conditions in Eq.~\ref{potential} are valid if the number of electrons near the grain surface is small compared to the total number of electrons in the grain,
i.~e. $R/\lambda_F \gg 1$, where $R$ is the grain radius and $\lambda_F$ is Fermi wavelength.
In Eq.~(\ref{h0}) $\psi^{\dag}(\textbf{r}) \, \, [\psi(\textbf{r})]$ is the creation [annihilation] operator and the integral is performed over the grain volume.

The second term in the r.~h.~s. of Eq.~(\ref{hamiltonian}) is responsible for Coulomb interaction
\begin{equation}
\label{int}
\hat{H}_{int}=\int d^3r \psi^{\dag}(\textbf{r})\psi^{\dag}(\textbf{r}')\frac{e^2}{|\textbf{r}-\textbf{r}'|}\psi(\textbf{r})\psi(\textbf{r}').
\end{equation}
\end{subequations}

The common method of studying the electronic properties of such systems is based on a mean-field approach. This approach reduces the many-body problem to a single particle problem with renormalized effective potential, where the effective potential contains all of the corrections due to the interaction between the electrons. However, in the present manuscript we use a different approach. Using the many-body Hamiltonian, Eq.~(\ref{hamiltonian}), we consider the Coulomb interaction, Eq.~(\ref{int}), perturbatively. This allows us to tackle the problem without reducing it to a single particle problem with renormalized potential $U(\textbf{r})$. We perform our calculations in the basis of the eigenfunctions of the non-interacting Hamiltonian $\hat{H}_0$, Eq.~(\ref{h0}) and take into account interaction effects, Eq.~(\ref{int}), summing the appropriate series of Feynman diagrams.

Our consideration is valid for metallic grains meaning that the grain itself is a good conductor. This means that the grain size $R$ is much larger than the Fermi wavelength $\lambda_F$. In addition, the disorder within the grain is weak such that electrons within the grain are delocalized. If these conditions are valid there is a small parameter in the problem - the invers grain conductance $1/g$, where $g$ is the grain conductance. The perturbation theory is built then on the expansion with respect to this small parameter.

To calculate the dynamic polarizability $\alpha(\omega)$ we study the electronic response of a single metallic grain to the external electric potential $\Phi^{ext}(\textbf{r})e^{-i\omega t}$. The polarizability is defined as a coefficient between the external electric field $\textbf{E}$ and the induced dipole moment, $\textbf{p}(\omega)=\alpha(\omega)\textbf{E}$. Thus, the problem is reduced to find the electric dipole moment $\textbf{p}(\omega)$ excited by the external field
\begin{equation}
\textbf{p}(\omega) =\int d^3 r \,  \textbf{r}\rho^{in}(\omega, \textbf{r}),
\end{equation}
where $\rho^{in}(\omega, \textbf{r})$ is the field-induced electronic charge density. Within the linear response theory $\rho^{in}(\omega, \textbf{r})$
is given by the following expression, see e.\,g. Ref.~\cite{Efetovbook}
\begin{equation}\label{rho_ind}
\rho^{in}(\omega, \textbf{r})=-\int d^3 r'\Phi(\omega, \textbf{r}')\chi(\omega, \textbf{r},\textbf{r}').
\end{equation}
Here $\chi(\omega, \textbf{r},\textbf{r}')$ is the electron density-density correlation function and $\Phi(\omega, \textbf{r})$ is the total electric potential, which is a sum of the external potential $\Phi^{ext}(\omega, \textbf{r})$ and the induced potential $\Phi^{in}(\omega, \textbf{r})$, $\Phi(\omega, \textbf{r})=\Phi^{ext}(\omega, \textbf{r})+\Phi^{in}(\omega, \textbf{r})$.
Equation~(\ref{rho_ind}) corresponds to the random phase approximation~\cite{Pines}.

In the quasi-static limit the potential $\Phi(\omega, \textbf{r})$ can be found by solving Poisson equation $\nabla^2\Phi(\omega, \textbf{r}) = -4 \pi \rho^{in}(\omega, \textbf{r})$. Using this expression in Eq.~(\ref{rho_ind}) we obtain the following integral equation for the potential
\begin{equation}
\label{phi_equat}
\nabla^2\Phi(\omega, \textbf{r}) = 4\pi\int d^3 r'\Phi(\omega, \textbf{r}')\chi(\omega, \textbf{r},\textbf{r}').
\end{equation}
In Eq.~(\ref{phi_equat}) the electron density-density correlation function $\chi(\omega,\textbf{r},\textbf{r}')$ is expressed in terms of electron Green's functions $G(\epsilon,\textbf{r}',\textbf{r})$ as
$\chi(\omega,\textbf{r},\textbf{r}')=-T\sum_{\epsilon}G(\epsilon+\omega,\textbf{r},\textbf{r}') G(\epsilon,\textbf{r}',\textbf{r})$.
Using the properties of Green's functions $\chi(\omega,\textbf{r},\textbf{r}')$ can be written in the form, see e.\,g. Ref.~\cite{Mahanbook}
\begin{equation}\label{chi}
\chi(\omega,\textbf{r},\textbf{r}')=2e^2\sum_{n,n'}\frac{f(\epsilon_n)-f(\epsilon_{n'})}{\omega-\epsilon_n+ \epsilon_n'} \psi^*_n(r)\psi_n(r')\psi^*_{n'}(r')\psi_{n'}(r),
\end{equation}
where $f(\epsilon_n)$ is the Fermi-Dirac distribution function with $\epsilon_n$ being the energy level of the quantum system and $\psi_n(r)$ is the eigenfunction of the non-interacting Hamiltonian with
$n$ being the full set of quantum numbers that describe the state. Below we consider the solution of Eq.~(\ref{phi_equat}) in details.

\section{Solution} To solve Eq.~(\ref{phi_equat}) we introduce the eigenfunctions $\psi_{nlm}$ and
corresponding eigenvalues $\epsilon_{ln}$ for electrons in the spherical potential well $U({\textbf r})$ of radius $R$
\begin{equation}
\psi_{nlm}= \beta_{ln} j_l(\alpha_{ln}r/R) Y_{lm}(\theta,\phi), \hspace{0.2cm}
\epsilon_{ln} = \alpha_{ln}^{2}/2mR^2,
\end{equation}
where $\alpha_{ln}$ is the $n$-th zero of the spherical Bessel function $j_l (x)$, $Y_{lm} (\theta,\phi)$ is the spherical harmonic, and $\beta_{ln} = (2^{1/2}/R^{3/2}) j'_l(\alpha_{ln})$ is the normalization coefficient with $j'_l(\alpha_{ln})$ being the derivative of the Bessel function taken at $x = \alpha_{ln}$.

The potential $\Phi(\omega, \textbf{r})$ in Eq.~(\ref{phi_equat}) consists of two parts: applied external potential and induced dipole part~\cite{Rice}.
Choosing the direction of the $z$ axis along the external electric field $E_{\omega}$ we obtain
\begin{equation}
\Phi(\omega, \textbf{r})=-E_{\omega}r\cos{\theta} + \Phi^{in}(\omega, \textbf{r}).
\end{equation}
Here the induced electric field $\Phi^{in}(\omega, \textbf{r})$ vanishes at infinity, $\lim_{|\textbf{r}|\rightarrow\infty}\nabla\Phi^{in}(\omega, \textbf{r})=0$.
In addition, the potential $\Phi^{in}(\omega, \textbf{r})$ and its derivative are continuous everywhere
including the grain surface. Outside the grain the charge density $\rho^{in}(\omega, \textbf{r})$ is zero therefore
the potential $\Phi^{in}(\omega, \textbf{r})$ must satisfy the Laplace's equation
\begin{equation}
\nabla^2 \Phi^{in}(\omega, \textbf{r})=0, ~|r|>R.
\end{equation}
Thus, the potential outside the grain can be found as a classical dipole potential
\begin{equation}
\Phi^{in}(\omega, \textbf{r})=\frac{c_0(\omega) E_{\omega}R^3\cos\theta}{r^2},
\end{equation}
where $c_0(\omega)$ is yet unknown function of frequency $\omega$.

Inside the grain one can seek the solution in the form
\begin{equation}
\Phi^{in}(\omega, \textbf{r}) = E_{\omega} \left[ c_0(\omega) r + R\,  f(\omega, r) \, \right]\cos\theta,
\end{equation}
where $f(\omega, r)$ is some unknown function of frequency $\omega$ and distance $r$. Introducing the function $g(\omega, r)$ as
\begin{equation}
\rho^{in}(\omega, \textbf{r})=\frac{E_{\omega}R}{4\pi}g(\omega, r)\cos\theta,
\end{equation}
we can rewrite Eq.~(\ref{rho_ind}) as follows
\begin{equation}
\label{int_equat}
g(\omega, r)=\int_0^R ds s^2 q(\omega, r,s)\left[ (c_0(\omega) -1)s/R + f(\omega, s) \right],
\end{equation}
with the kernel $q(\omega, r,s)$ being
\begin{equation}
\frac{q(\omega, r,s)}{4\pi} = \int \cos\theta d\phi d\theta\int \cos\theta' d\phi' d\theta' Y_{10}(\theta)Y_{10}(\theta') \chi(\omega, r, s).
\end{equation}

Since we are interested in the dipole response of the grain we expand the functions $f(\omega, r)$ and $g(\omega, r)$
in Eq.~(\ref{int_equat}) using the orthogonal set $\phi_n(r)=j_1(\alpha_{1n}r/R)$
\begin{eqnarray}
\label{fg}
f(\omega, r)&=& \sum_{n=1}^{\infty} c_n(\omega) j_1\left(\frac{\alpha_{1n}r}{R}\right), \nonumber \\
g(\omega, r)&=& \sum_{n=1}^{\infty} c_n(\omega) \left(\frac{\alpha_{1n}}{R}\right)^2j_1\left(\frac{\alpha_{1n}r}{R}\right).
\end{eqnarray}
Thus, the problem of calculating the induced potential $\Phi^{in}(\omega, \textbf{r})$ inside the grain
is reduced to the problem of calculating the frequency dependent coefficients $c_n(\omega)$ in Eq.~(\ref{fg}). Using these coefficients
the dynamic polarizability $\alpha(\omega)$ of a grain can be written as follows
\begin{equation}
\label{alpha}
\alpha(\omega) = c_0(\omega) R^3,
\end{equation}
with $R^3 \equiv \alpha_{cl}(0)$ being the static classical polarizability. The
coefficient $c_0(\omega)$ can be considered as a dimensionless dynamic polarizability and
can be found using the boundary conditions
\begin{equation}
\label{c0}
c_0(\omega) = -\frac{1}{3}\sum_{n=1}^{\infty}c_n(\omega) j_1'(\alpha_{1n})\alpha_{1n}.
\end{equation}
To calculate the frequency dependent coefficients $c_n(\omega)$ in Eq.~(\ref{c0}) we
substitute Eqs.~(\ref{fg}) into Eq.~(\ref{int_equat}) to obtain the system
of linear equations for coefficients $c_n (\omega)$
\begin{subequations}
\label{aA}
\begin{equation}
A_{mn}(\omega) \, c_n (\omega) = a_m (\omega),
\end{equation}
where
\begin{eqnarray}
\label{coefficients}
a_m (\omega') =\frac{3}{2\sqrt{2}\pi}\sqrt{\frac{\lambda_F}{a_B}}\frac{\lambda_F}{R}\sum_{l,n_1,n_2} \left( \frac{f_{ln_1}-f_{(l-1)n_2}}{\omega'-\epsilon_{ln_1}+\epsilon_{(l-1)n_2}}\frac{l}{\gamma_{ln_1}\gamma_{(l-1)n_2}} B^1_m(l,n_1,n_2)C^1(l,n_1,n_2)\right.\nonumber \\+
\left.
\frac{f_{ln_1}-f_{(l+1)n_2}}{\omega'-\epsilon_{ln_1}+\epsilon_{(l+1)n_2}}\frac{l+1}{\gamma_{ln_1}\gamma_{(l+1)n_2}} B^2_m(l,n_1,n_2)C^2(l,n_1,n_2)\right),
\end{eqnarray}
\begin{eqnarray}
A_{mn} (\omega') = \frac{3}{2\sqrt{2}\pi}\sqrt{\frac{\lambda_F}{a_B}}\frac{\lambda_F}{R}\sum_{l,n_1,n_2} \left( \frac{f_{ln_1}-f_{(l-1)n_2}}{\omega'-\epsilon_{ln_1}+\epsilon_{(l-1)n_2}}\frac{l}{\gamma_{ln_1}\gamma_{(l-1)n_2}} B^1_m(l,n_1,n_2)B^1_n(l,n_1,n_2) \right.\nonumber \\ \left.+ \frac{f_{ln_1}-f_{(l+1)n_2}}{\omega'-\epsilon_{ln_1}+\epsilon_{(l+1)n_2}}\frac{l+1}{\gamma_{ln_1}\gamma_{(l+1)n_2}} B^2_m(l,n_1,n_2)B^2_n(l,n_1,n_2)\right)\nonumber \\
-\frac{1}{3}a_m(\omega')\alpha_{1n}j_1'(\alpha_{1n})-\delta_{mn}\alpha_{1n}^2\gamma_{1n}.
\label{coefficients2}
\end{eqnarray}
\end{subequations}
Here $a_B=1/me^2$ is the Bohr radius and we introduce the notation $\gamma_{ln} = \frac{1}{2}\left[j_l'(\alpha_{ln})\right]^2$, and
the dimensionless frequency $\omega' = \omega/(\omega_p/\sqrt{3})$, with $\omega_p^2 = (4/3)\pi\nu e^2 v_F^2$ being the plasma frequency.

The quantities $B^1_m(l,n_1,n_2)$, $B^2_m(l,n_1,n_2)$, $C^1(l,n_1,n_2)$, and $C^2(l,n_1,n_2)$ in Eqs.~(\ref{coefficients})
and (\ref{coefficients2}) are defined as
\begin{eqnarray}
B^{1(2)}_m(l,n_1,n_2) &=& \int_{0}^{1}dx x^2 j_1(\alpha_{1n}x)j_l(\alpha_{ln_1}x)j_{l\mp1}(\alpha_{(l\mp1)n_2}x), \nonumber \\
C^{1(2)}(l,n_1,n_2) &=& \int_{0}^{1}dx x^3 j_l(\alpha_{ln_1}x)j_{l\mp1}(\alpha_{(l\mp1)n_2}x),
\end{eqnarray}
where the upper index $1$ $(2)$ corresponds to $-$ $(+)$.
The Fermi-Dirac functions $f_{ln}$ in Eqs.~(\ref{coefficients})
and (\ref{coefficients2}) are taken at zero temperature and defined as
\begin{equation}
\label{destribution}
f_{ln}= \left\{
\begin{array}{lr}
1, \hspace{0.3cm} \alpha_{ln} < \alpha_F \\
0, \hspace{0.3cm} \alpha_{ln} > \alpha_F,
\end{array}
\right.
\end{equation}
where the dimensionless parameter $\alpha_F$ defines the
Fermi energy $E_F=(\hbar^2/2mR^2)\alpha_F^2$. Using Eq.~(\ref{destribution}) for the distribution function $f_{ln}$ one can calculate the number of the electrons
in the conducting band as follows
\begin{equation}
N=2\sum_{ln}(2l+1)f_{ln}.
\end{equation}

For numerical calculations we used the complex frequency $\omega'+i\gamma'$ with $\gamma'=0.006$ being the dimensionless damping factor. The damping factor $\gamma'$ was introduced to avoid infinite values of the polarizability at the resonances and
does not correspond to the real plasmon resonance broadening. The analysis of the resonance broadening is beyond the scope of the present work.

Using Eqs.~(\ref{alpha}) - (\ref{coefficients2}) we numerically calculate the dynamic polarizability $\alpha(\omega)$ in Eq.~(\ref{alpha}) for grains of
different sizes. The final results are shown in Fig.~\ref{fig50} and Fig.~\ref{fig25}

\section{Conclusions} We studied the dynamic polarizability of spherical metallic grains using quantum mechanical treatment. We numerically investigated the frequency behavior of polarizability for relatively large grains. We showed that the main resonance peak corresponding to the surface plasmon mode is blue-shifted and some minor secondary resonances above and below the main peak exist. We studied the dependence of blue shift as a function of grain size and compared our results with the classical polarizability.

\acknowledgements{ We thank G.~Weick for helpful discussions. This research was supported by an award from Research Corporation for Science Advancement.}

\end{document}